\documentclass{ws-p9-75x6-50}

\begin{document}

\newcommand{\approxgt}{\mathrel{\hbox{\rlap{\lower.55ex \hbox {$\sim$}}
        \kern-.3em \raise.4ex \hbox{$>$}}}}

\def\approxlt{\mathrel{\hbox{\rlap{\lower.55ex \hbox {$\sim$}}
        \kern-.3em \raise.4ex \hbox{$<$}}}}

\title{ X-rays from the environment of supermassive black holes in active galaxies}

\author{Stefanie~~Komossa}

\address{Max-Planck-Institut fuer extraterrestrische Physik, Postfach 1312, D-85741 Garching,
Germany  \\ E-mail: skomossa@xray.mpe.mpg.de}


\maketitle

\small 
\vspace*{-7.1cm}
\begin{verbatim}
    Invited review, IX. Marcel Grossmann Meeting on General Relativity,
                    Gravitation and Relativistic Field Theories
    Session: ``Astrophysics of Neutron Stars and Black Holes: Observations''
    (Rome, July 2000); 2-page condensate will be published in World Scientific
    extended (10 pages) version will appear in the online proceedings 
    at  {http://www.icra.it/MG/mg9/mg9.htm}
    \end{verbatim}
    \vspace*{4.5cm}

\normalsize

\abstracts{
X-rays are a powerful probe of the physical
conditions in the nuclei of active galaxies (AGN).
We review the 
X-ray properties of radio-quiet AGN, LINERs and ULIRGs
based on observations carried out with the  
X-ray satellite {\sl ROSAT}.
We then summarize 
the observations  
of giant X-ray flares 
from non-active galaxies, interpreted as stellar tidal disruptions by
supermassive black holes. 
}

\section{Introduction}

Active galaxies (AGN) are believed to be powered by accretion
onto supermassive black holes at their centers.
X-ray observations currently provide the most powerful
way to explore the black hole region of 
AGN. The bulk energy output of active galactic nuclei
is in the X-ray bandpass, and the X-ray properties of AGN, particularly
their high luminosity and rapid variability, originally gave the
best evidence that supermassive black holes (SMBHs hereafter) do reside
at the centers of AGN.
  
Below, we first summarize the results of X-ray observations of radio-quiet 
AGN, LINERs and ultraluminous IR galaxies carried
out with the {\sl ROSAT} (Tr\"umper 1983) X-ray observatory{\footnote{{\sl ASCA}
results on AGN are presented by K. Iwasawa, these proceedings}}. 
We then address 
the X-ray search for SMBHs in {\em non}-active galaxies.

\section{Active Galactic Nuclei (AGN)}


In the course of the {\sl ROSAT} all-sky survey (Voges et al. 1996), many new AGN have
been identified{\footnote{e.g.,
Schwope et al. 2000, Wei et al. 1999, Appenzeller et al. 1998, Bade et al. 1995, 1998}}, 
confirming that X-ray surveys 
are very efficient in finding  
new AGN, and many have been studied in detail in the later phase
of pointed observations on selected bright targets.  
Large-sample studies of radio-quiet quasars gave a mean powerlaw photon 
index $\Gamma_{\rm x} \simeq -2.5$ and indicated a flattening of the
X-ray spectrum with increasing energy/redshift, interpreted as the onset of a  
new spectral component (e.g.,  
Schartel et al. 1996a,b, Ciliegi \& Maccacaro 1996, Yuan et al. 1998).  
A large sample study of Seyfert galaxies observed during
the {\sl ROSAT} all-sky survey showed the spetra to span
a range in powerlaw indices between $\Gamma_{\rm x}=-1.6...-3.4$ (Walter \& Fink 1993). 
Many spectra of Seyfert galaxies show complexity in form of soft 
excesses or ionized absorbers.


X-rays at the centers of AGN originate in the 
accretion-disk region (see Mushotzky et al. 1993, Collin et al. 2000 for reviews). 
On larger scales, but still within the nucleus, X-rays might be emitted
by a hot intercloud medium at distances of the broad or narrow-line region,
BLR or NLR (e.g., Elvis et al. 1990; Ogle et al. 2000).   

The X-rays which originate from the accretion-disk region
are reprocessed in form of absorption and partial re-emission 
(e.g., Netzer 1993, 1996, Krolik \& Kriss 1995) 
as they
make their way out of the nucleus. 
The reprocessing bears the disadvantage of veiling the {\em intrinsic}
X-ray spectral shape,
and the spectral disentanglement of many different potentially contributing
components is not always easy. 
However, reprocessing also offers the unique chance 
to study the physical conditions
and dynamical states of the reprocessing material, like: the ionized absorber;
the torus, which plays an important role in AGN unification schemes 
(Antonucci 1993, see also Elvis 2000); and also the BLR and NLR.
Detailed modeling of the reprocessor(s) is also necessary
to recover the shape and properties of the {\em intrinsic} X-ray spectrum.  
In the soft X-ray bandpass, the best-studied
reprocessor so far is the so-called `warm absorber', highly ionized
gas in the circum-nuclear environment.   
This is because it is nearly  completely ionized in Hydrogen
and Helium, and does not absorb
at very soft energies, like torus, BLR and NLR will, if they
are located along the line-of-sight.  

With {\sl ROSAT}, the signatures of a warm absorber, absorption edges
of highly ionized oxygen ions at 
$E_{\rm OVII}=0.74$ keV and $E_{\rm OVII}=0.87$ keV, were first detected in MCG-6-30-15
(Nandra \& Pounds 1992). Detailed studies of many
other AGN followed, with  
the following results (see Komossa 1999 for a review on warm absorbers): 

Ionized absorbers are observed in $\sim$50\% of the well-studied 
Seyfert galaxies{\footnote{e.g., MCG-6-30-15, 
Nandra \& Pounds 1992; NGC\,3783, Turner et al. 1993;
NGC\,5548, Nandra et al. 1993; NGC\,4051, Pounds et al. 1994; NGC\,3227, 
Komossa \& Fink 1997b; a more complete list of objects including observations
from other X-ray satellites is given by Komossa (1999)}}.
They are less abundant in quasars (e.g.,  
Ulrich \& Molendi 1996, Laor et al. 1997), but do occur 
in some{\footnote{e.g., 3C351, Fiore et al. 1993; 3C212, Mathur 1994;
MR\,2251-178, Komossa 2001}}.  
In many Seyfert galaxies,  warm absorbers replaced the soft excess,
i.e., steep X-ray spectra in the soft X-ray band, originally
thought to stem from black-body-like soft excess emission 
turned out to be caused by the presence of warm absorbers. 

From the general lack of rapid variability
in response to intrinsic luminosity variations,
low densities of the absorbers were generally inferred.
For example, the limits on density and location
of the ionized material in NGC\,4051 are $n_{\rm{H}} \approxlt$ $3 \times 10^{7}$cm$^{-3}$
and $r \approxgt 3 \times 10^{16}$ cm,  
its column density is $N_{\rm w} \simeq 10^{22.7}$ cm$^{-2}$, and its
electron temperature is a few 10$^5$ K (Komossa \& Fink 1997a).
Several approaches were made to assess the thermal
stability of the warm absorber in the context of 
multi-phase cloud equilibrium models (e.g. Reynolds \& Fabian 1995,
Komossa \& Fink 1997a,b), addressing the
possibility that the ionized material is in equilibrium
with a hotter (or cooler) gas phase. Results are summarized
in Fig. 1.  

\begin{figure}[t]
\begin{center}
\epsfxsize=20pc 
\epsfbox{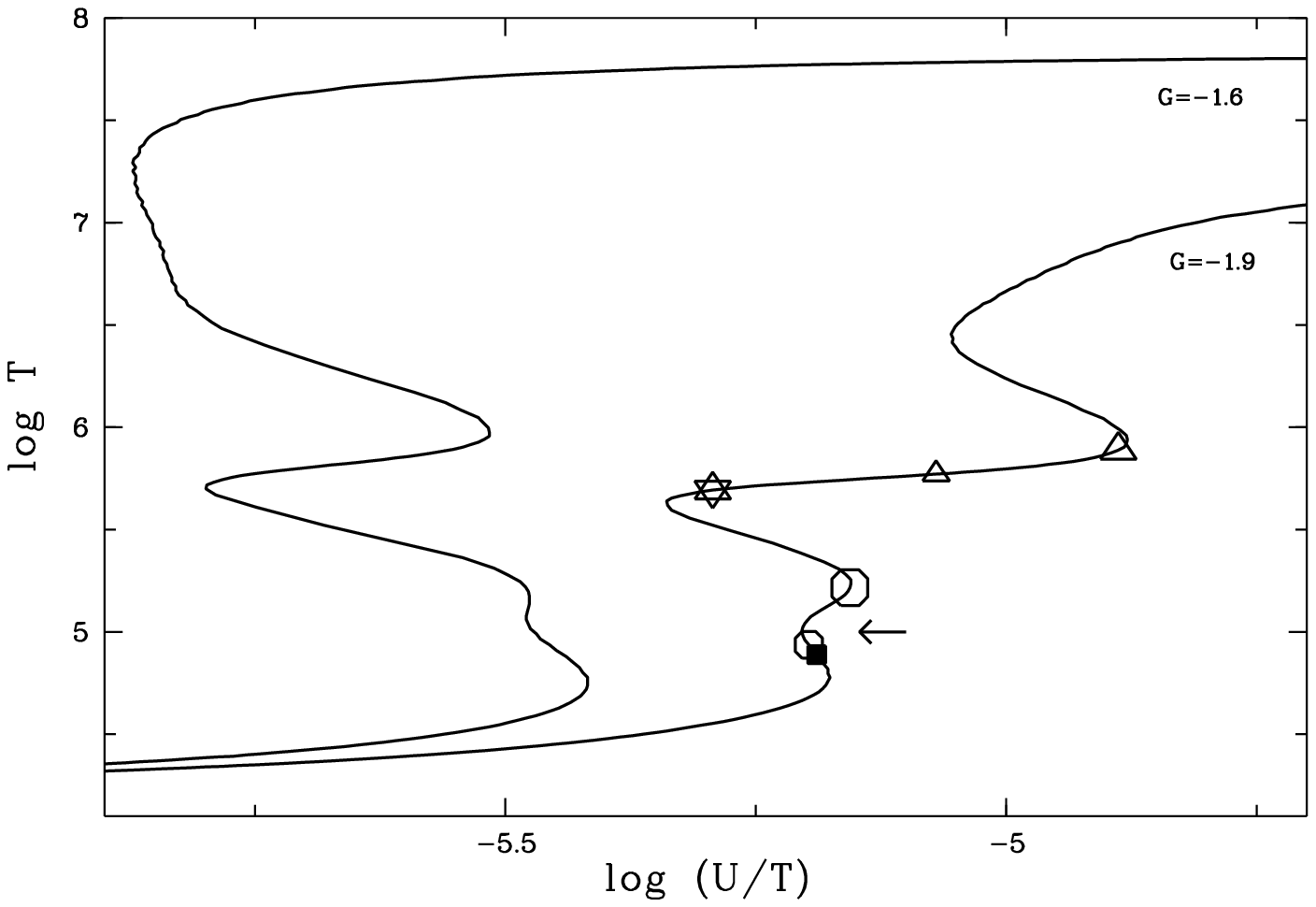} 
\epsfxsize=20pc

\vspace*{0.1cm}

\epsfbox{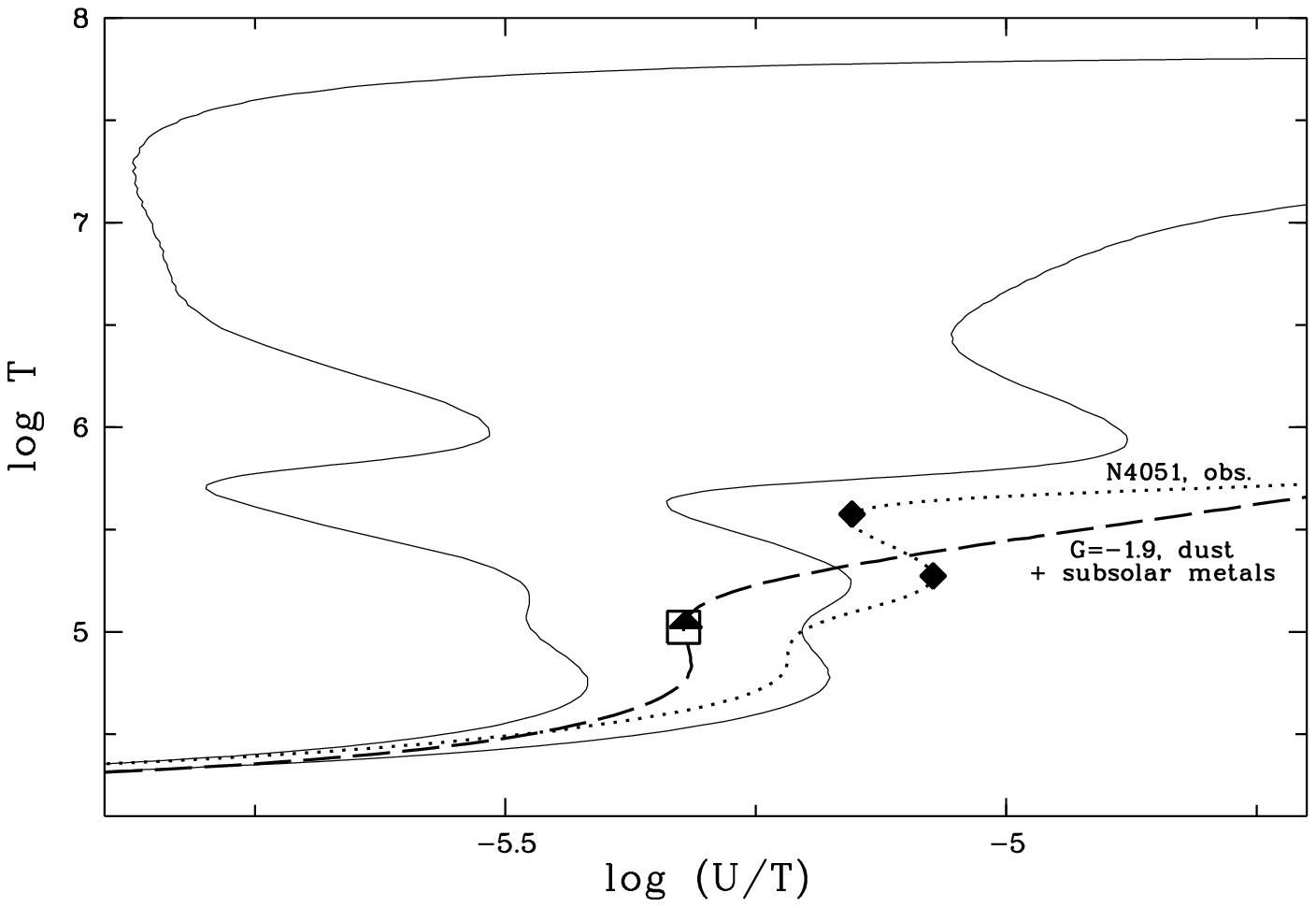}    
\end{center} 
\caption{Equilibrium gas temperature $T$ versus pressure ($U/T$, where $U$ is
the ionization parameter), and locations of selected warm absorbers.
Regions where $T$ is multi-valued for constant $U/T$ and
where the gradient of
the equilibrium curve is positive  allow for the co-existence
of several phases in pressure balance.
The equilibrium curves are shown for different ionizing continua
and gas properties, and the locations of several warm absorbers 
are marked. {\bf Upper panel:} 
The solid lines correspond to spectral energy distributions (SEDs)
with (i) $\Gamma_{\rm x}$ = --1.6 (left)
and (ii) $\Gamma_{\rm x}$ = --1.9 (right), and $\alpha_{\rm uv-x}$ = --1.4. 
The symbols mark the locations of warm absorbers, derived from X-ray spectral
fits. Filled square: Mrk\,1298; star: RXJ\,0134-4258; 
arrow: MCG-6-30-15 (Reynolds \& Fabian 1995); triangles: PG\,1404+226 high- and low-state;
circles: PKS\,2351-154 high- and low-state. 
{\bf Lower panel:} curves (i) and (ii) from the upper panel
are re-drawn. The dashed line corresponds to an input SED
with $\Gamma_{\rm x}$ = --1.9 but instead of solar gas abundances we adopted
depleted metal abundances and mixed dust with the gas (see Komossa \& Fink 1997b for details);
the dotted line corresponds to the observed SED of NGC 4051 (Komossa \& Fink 1997a)
with $\Gamma_{\rm x}$ = --2.3. 
Filled triangle: location of the dusty warm absorber of NGC\,3227; open square:
dusty warm absorber of NGC\,3786, lozenges: NGC\,4051 during the Nov. 1993 (upper lozenge)
and Nov. 1991 (lower lozenge) observation. } 
\end{figure}

Whereas the existence of warm absorbers was first deduced
from X-ray spectra, there is now evidence that the same material
also shows up in other spectral bands, and the multi-wavelength
approach is a powerful one.
There is good evidence that the dust which causes the
optical reddening of several AGN is mixed with the
ionized absorber (Komossa \& Bade 1998 and references therein;  
see Tab. 1 of Komossa 1999 for a list 
of {\em dusty} warm absorbers). Further, several AGN have
been presented where UV- and X-ray warm absorber are likely
one and the same component, or are related to each other
(e.g., Mathur 1997 and references therein). 

Finally, warm absorbers have been invoked to explain some
otherwise puzzling observations like the dramatic spectral
variability of the Narrow-line Seyfert\,1 galaxy 
RXJ0134-4258 (Komossa \& Meerschweinchen 2000), 
or the apparent huge excess {\em cold} X-ray absorption
in high-redshift quasars (Schartel et al. 1997, Fabian et al. 2001).  	 

Absorption features at $\sim$1.1 keV (e.g., Ulrich \& Molendi 1996)
have been interpreted in terms of relativistic
outflow of the ionized medium (Leighly et al. 1997, see also Brandt et al. 1994),
but several alternative explanations -- most of them still linked
to a warm absorber albeit in a different way -- have been
suggested (e.g., Ulrich et al. 1999, Nicastro et al. 1999, Turner et al. 1999).

Recent {\sl Chandra} observations of the Seyfert galaxies
NGC\,5548 and NGC\,3783 revealed a rich absorption
line spectrum (Kaastra et al. 2000, Kaspi et al. 2000) and demonstrate
the power of high-resolution spectroscopy to probe the physics  
of the ionized gas in the nuclei of AGN.

In summary, the study of the ionized absorbers provides a wealth of information
about the nature of the warm absorber itself, its relation
to other components of the active nucleus, and the intrinsic
AGN X-ray spectral shape, and leads to a
 better  understanding of the black hole region of AGN, and its
cosmic evolution.

\section{LINER galaxies}

LINER (Low-Ionization Nuclear Emission-Line Region)  galaxies
are characterized by their optical emission line spectrum
which shows a lower degree of ionization 
than AGN (Heckman 1980). Their major power source and line excitation mechanism
have been a subject of lively debate ever since their discovery.
LINERs manifest the most common type of activity in the local universe.
If powered by accretion, they probably represent the low-luminosity end
of the quasar phenomenon, 
and their presence has relevance to, e.g., the evolution of quasars,
the faint end of the Seyfert luminosity function, the soft X-ray background,
and the presence of SMBHs in nearby galaxies.

The soft X-ray properties of LINERs are inhomogeneous
(Komossa et al. 1999, and references therein). Whereas 
the spectra of about
50\% of them are best described by AGN-like powerlaws, the others 
are dominated by thermal Raymond-Smith emission 
with evidence that the spectra are more complex than
emission from single-temperature gas in collisional-ionization
equilibrium.
X-ray luminosities are in the range $\sim$10$^{38-41}$ erg/s
(for comparison:  Seyfert\,1 galaxies and quasars typically exceed 10$^{42}$ erg/s).  
The general absence of short-time scale (hours-weeks) variability    
is consistent with the suggestion that LINERs accrete in the
advection-dominated mode (e.g, Yi \& Boughn 1998, 1999, and references therein).  

\begin{figure}[t]
\begin{center}
\epsfxsize=21pc 
\epsfbox{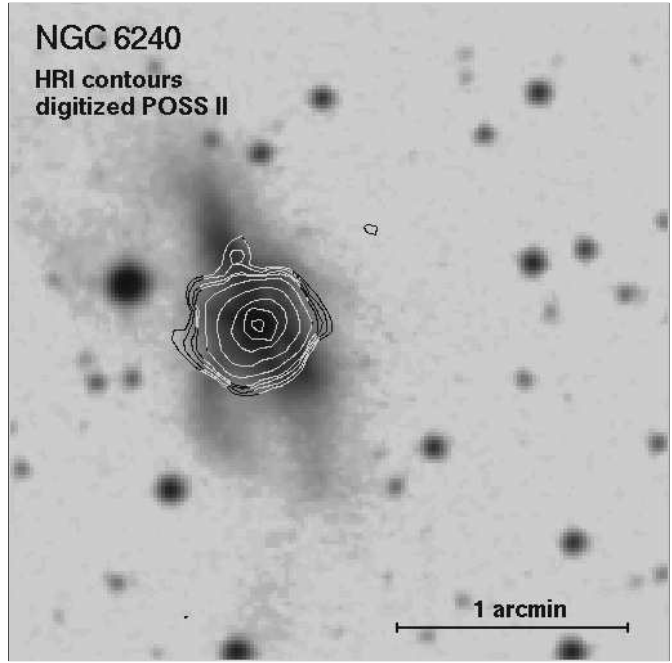} 
\end{center}
\caption{ {\sl ROSAT} HRI X-ray contours of the ultraluminous IR galaxy
NGC\,6240, showing the presence of luminous
extended emission, overlaid on an optical image of
the galaxy (taken from Komossa et al. 1998). }
\end{figure}

\section{Ultraluminous Infrared galaxies (ULIRGs)} 

ULIRGs, characterized by their huge power-output in the infrared 
which exceeds 10$^{12}\,L_\odot$ (Sanders \& Mirabel 1996),
are powered by massive starbursts or SMBHs. The discussion,
which one actually dominates received a lot
of attention in recent years (e.g., Joseph 1999, Sanders 1999).

Many distant SCUBA sources, massive and dusty galaxies,
are ULIRG equivalents at high redshift.
{\em Local} ULIRGs 
are ideal laboratories to study the physics of galaxy formation
and the processes of IGM enrichment (e.g., Heckman 1999), and the physics of 
superwinds (Breit\-schwerdt \& Komossa 2000 and references therein)
driven by the nuclear starbursts. 

With a redshift $z=0.024$ and a far-infrared
luminosity of $\sim 10^{12} L_{\odot}$,
NGC\,6240 is one of the nearest members of the class of ULIRGs.
Whereas X-rays from distant Hyperluminous IR galaxies, HyLIRGs,
were not detected by Wilman et al. (1999),
and the ULIRGs in the study of Rigopoulou et al. (1996)
were X-ray weak, NGC\,6240 turned out to be exceptionally
X-ray luminous. It is 
the most luminous emitter in extended soft X-rays among 
ULIRGs (Komossa et al. 1998).
Starburst-driven superwinds are the most likely interpretation
of the extended emission (see Schulz \& Komossa 1999 for alternatives),
albeit being pushed to their limits to explain the huge
power output (Schulz et al. 1998).

\newpage

\section{X-ray search for SMBHs in non-active galaxies}

Do SMBHs exist at the centers of {\em all} galaxies ? 
This question is relevant for, e.g., the study of 
the formation and evolution of galaxies and
AGN in general (see Kormendy \& Richstone 1995, Schulz \& Komossa 1999 for reviews).  

\begin{table*}[b]
\begin{center}
\footnotesize
\caption{Summary of the X-ray properties of NGC\,5905 and RX\,J1242--1119
   during outburst. $L_{\rm x}$ gives the intrinsic luminosity in the
   (0.1--2.4) keV band using $H_0 = 50$ km/s/Mpc, 
    $T_{\rm bb}$ is the black body
    temperature derived from a black body fit to the data. }
\begin{tabular}{ccccl}
  \hline
  \noalign{\smallskip}
name & redshift & $kT_{\rm bb}$ [keV] & $L_{\rm x,bb}$ [erg/s] & references \\
  \noalign{\smallskip}
  \hline
  \hline
  \noalign{\smallskip}
NGC\,5905 & 0.011 & 0.06$\pm{0.01}$ & 3\,10$^{42}$$^*$ & Bade et al. 1996, Komossa \& Bade 1999 \\
  \noalign{\smallskip}
RXJ1242$-$11 & 0.050 & 0.06$\pm{0.01}$ & 9\,10$^{43}$~ & Komossa \& Greiner 1999 \\
  \noalign{\smallskip}
\hline
  \noalign{\smallskip}
  \noalign{\smallskip}
\end{tabular}
\end{center}

  \noindent{\footnotesize $^{*}$ Mean luminosity during the outburst; since the flux
 varied by a factor $\sim$3 during the observation, the peak luminosity is higher.}
\end{table*}

One efficient method to search for such usually quiescent 
SMBHs in nearby, {\em non}-active galaxies
is to make use of the expected flares from tidally disrupted stars
(e.g., Lidskii \& Ozernoi 1979, Rees 1988, 1989), and the
first excellent candidates for tidal disruption events
in {\em optically non-active} galaxies have been discovered
in the last few years (e.g., Bade et al. 1996, Komossa \& Bade 1999). 
Depending on its orbits, a star approaching
a SMBH will be tidally disrupted
if \begin{equation}
r_{\rm t} \approx r_* ({M_{\rm BH}\over M_*})^{1 \over 3} ~~,
\end{equation}
where $r_{\rm t}$ is the tidal radius, and $M_{\rm BH}$ the black hole
mass. 
The star is first heavily distorted, then disrupted.
About half of the gaseous debris will be unbound and gets
lost from the system (Young et al. 1977).
The rest will be eventually accreted by the black hole
(Cannizzo et al. 1990, Loeb \& Ulmer 1997).
The debris, first spread over a number of orbits,
quickly circularizes (Rees 1988, Cannizzo et al. 1990)
due to the action of strong
shocks when the most tightly bound debris interacts with
other parts of the stream (Kim et al. 1999).
The process is accompanied by
a flare of electromagnetic radiation with the maximum in the 
UV or EUV spectral region, and with a duration of months to years.     

\begin{figure}[ht]
\begin{center}
\epsfxsize=18pc 
\epsfbox{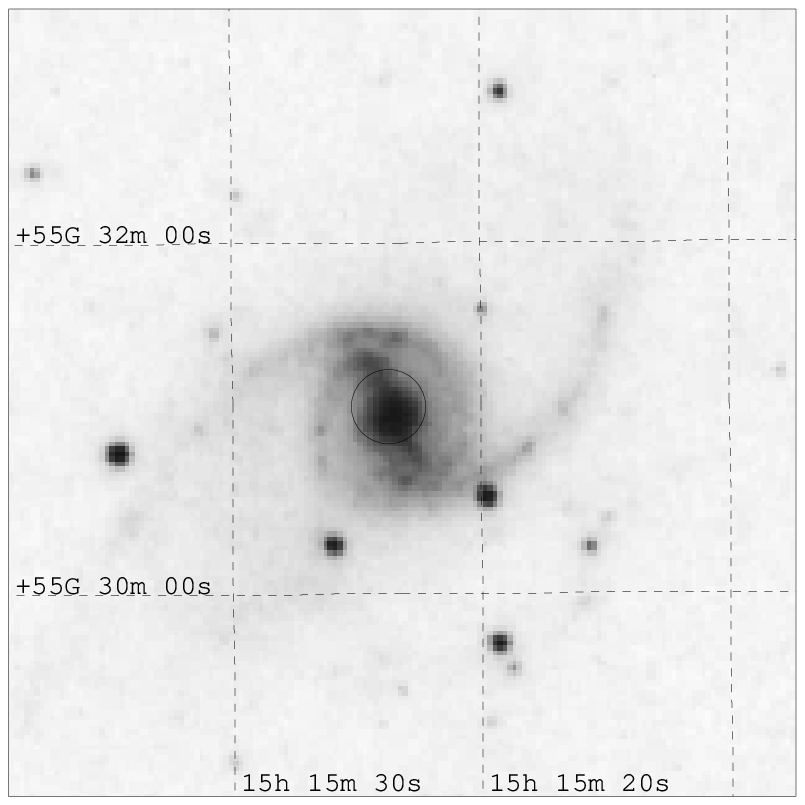} 
\end{center}
\caption{Optical image of NGC\,5905 and error circle of the X-ray outburst emission
(taken from Bade et al. 1996).}

\vspace*{0.7cm}

\begin{center}
\epsfxsize=26pc 
\epsfbox{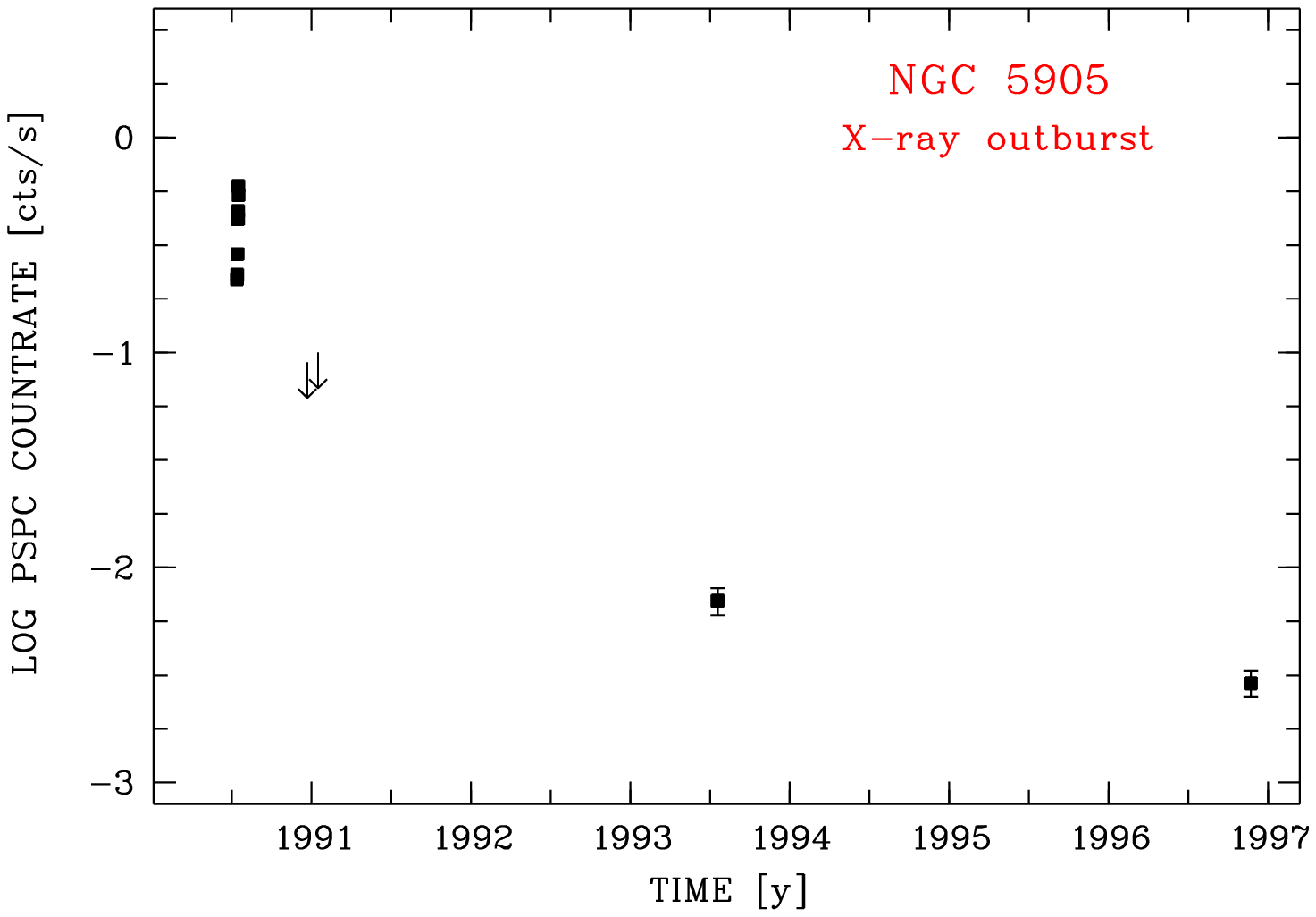} 
\end{center}
\caption{Long-term X-ray lightcurve of NGC\,5905 (taken from Komossa \& Bade 1999).} 
\end{figure}

The giant-amplitude X-ray outbursts which have recently been
discovered with {\sl ROSAT} from several non-active galaxies 
were interpreted in terms of such tidal disruption events (e.g., Bade et al. 1996, 
Komossa \& Bade 1999, Komossa \& Greiner 1999).  
The X-ray emission of these
galaxies varied by up to a factor $\sim$200 (NGC\,5905) and
they reached X-ray luminosities as large as 10$^{44}$ erg/s (RXJ1242-1119);
a summary of the observations is given in Table 1.  

These flares enable us to investigate the very vicinity of the SMBH.
Increased sensitivity of future X-ray instruments will allow to follow the disruption
and accretion process in detail, and study the strongly relativistic effects
related to them (e.g., they depend on relativistic precession effects around the Kerr metric).

\section*{Acknowledgments}
The {\sl ROSAT} project has been supported by the German 
Bundes\-mini\-ste\-rium
f\"ur Bildung, Wissenschaft, Forschung und Technologie
(BMBF/DLR) and 
the Max-Planck-Society. 
Preprints of this and related papers can be retrieved at 
http://www.xray.mpe.mpg.de/$\sim$skomossa/

\end{document}